\documentclass[11pt]{article}
\usepackage[dvips]{color}
\usepackage{epsfig}
\usepackage{amsmath}
\usepackage{graphicx}
\date{}
\begin{document}
\title{{\bf Ho\v{r}ava-Lifshitz scalar field cosmology: classical and quantum viewpoints}}
\author{Fatimah Tavakoli$^1$\thanks{%
e-mail: tavakoli09@gmail.com}\,\,, Babak Vakili$^1$\thanks{%
e-mail: b.vakili@iauctb.ac.ir}\,\, and  Hossein Ardehali$^{2}$\thanks{%
e-mail: hosseinardehali@gmail.com}
\\\\
$^1${\small {\it Department of Physics, Central Tehran Branch,
Islamic Azad University, Tehran, Iran}}\\
$^2${\small {\it Department of Physics, Science and Research Branch,
Islamic Azad University, Tehran, Iran}}} \maketitle

\begin{abstract}
In this paper, we study a projectable Ho\v{r}ava-Lifshitz cosmology
without the detailed balance condition minimally coupled to a
non-linear self-coupling scalar field. In the minisuperspace
framework, the super Hamiltonian of the presented model is
constructed by means of which, some classical solutions for scale
factor and scalar field are obtained. Since these solutions exhibit
various types of singularities, we came up with the quantization of
the model in the context of the Wheeler-DeWitt approach of quantum
cosmology. The resulting quantum wave functions are then used to
investigate the possibility of the avoidance of classical
singularities due to quantum effects which show themselves important
near these singularities.
\vspace{5mm}\noindent\vspace{0.8mm}\newline
\end{abstract}

\section{Introduction}
In recent years, a new gravitation theory based on anisotropic
scaling of the space $\bf{x}$ and time $t$ presented by Ho\v{r}ava.
Since the methods used in this gravitation theory are similar to the
Lifshitz work on the second-order phase transition in solid state
physics, it is commonly called Ho\v{r}ava-Lifshitz (HL) theory of
gravity \cite{Horava1,Horava2,Horava3,Horava4}. In general, HL
gravity is a generalization of general relativity (GR) at high
energy ultraviolet (UV) regime and reduces to standard GR in the low
energy infra-red (IR) limit. However, unlike another candidates of
quantum gravity the issue of the Lorentz symmetry breaking at high
energies, is described somehow in a different way. Here, the
well-known phenomenon of Lorentz symmetry breaking will be expressed
by a Lifshitz-like process in solid state physics. This is based on
the anisotropic scaling between space and time as

\begin{equation}\label{In1}
t\rightarrow b^z t,\hspace{0.5cm}{\bf x}\rightarrow b{\bf x},
\end{equation}where $b$ is a scaling parameter and $z$ is dynamical critical
exponent. It is clear that $z=1$ corresponds to the standard
relativistic scale invariance with Lorentz symmetry. Indeed, with
$z=1$, the theory falls within its IR limit. However, different
values of $z$ correspond to different theories, for instance what is
proposed in \cite{Horava1,Horava2} as the UV gravitational theory
requires $z=3$. In order to better represent the asymmetry of space
and time in HL theory, we write the space-time metric is its ADM
form, that is,
\begin{equation}\label{In2}
g_{\mu \nu}(t,{\bf x})=\left(%
\begin{array}{cc}
-N^2(t,{\bf x})+N_a(t,{\bf x})N^a(t,{\bf x}) & N_b(t,{\bf x}) \\
N_a(t,{\bf x}) & h_{ab}(t,{\bf x}) \\
\end{array}%
\right),
\end{equation}where $N(t,{\bf x})$ is the lapse function, $N^a(t,{\bf
x})$ are the components of the shift vector and $h_{ab}(t,{\bf x})$
is the spacial metric. There are two classes of HL theories
depending on whether the lapse function is a function only of $t$,
for which the theory is called projectable, or of $(t,{\bf x})$, for
which we have a non-projectable theory. Since in cosmological
settings the lapse function usually is chosen only as a function of
time, the corresponding HL cosmologies are projectable
\cite{vakili&kord}-\cite{Ardehali&Pedram}. In more general cases
however, one may consider the lapse function as a function of both
$t$ and ${\bf x}$ to get a non-projectable theory, see
\cite{non-projectable1,non-projectable2}. At first glance, it may
seem that imposing the lapse function to be just a function of time,
is a serious restriction. However, it should be noted that in the
framework of this assumption the classical Hamiltonian constraint is
no longer a local constraint but its integral over all spatial
coordinates should be done, which means that we have not a local
energy conservation. In \cite{Muko}, it is shown that this procedure
yields classical solutions to the IR limit of HL gravity which are
equivalent to Friedmann equations with an additional term of the
cold dark matter type. On the other hand, in homogeneous models like
Robertson-Walker metric, such spatial integrals are simply the
spatial volume of the space and thus the above mentioned dark dust
constant must vanish \cite{Sot}. In summary, it is worth to note
that although almost all physically important solutions of the
Einstein equations like Schwarzschild, Reissner-Nordstr\"{o}m, Kerr
and Friedmann-Lemaitre-Robertson-Walker space times, can be cast
into the projectable form by suitable choice of coordinates, most of
the results, in principle, may be extended to the non-projectable
case through a challenging but straightforward calculation
\cite{wein}.

Another thing to note about the HL theory is the form of its action
which in general consists of two kinetic and potential terms. Its
kinetic term ${\cal S}_K$, is nothing other than what comes from the
Einstein-Hilbert action. The form for the potential term is ${\cal
S}_V=\int d^4x\sqrt{-g}V[h_{ab}],$ where $V$ is a scalar function
depends only on the spacial metric $h_{ab}$ and its spacial
derivatives. Among the very different possible combinations that can
be constructed using the three-metric scalars \cite{vakili&kord},
Ho\v{r}ava considered a special form in $z=3$ theory known as
"detailed balance condition", in which the potential is a
combination of the terms $\nabla_a R_{bc} \nabla^a
R^{bc},\hspace{0.2cm}\nabla_a R_{bc}\nabla^b
R^{ac},\hspace{0.2cm}\nabla_a R \nabla^a R$, \cite{Horava2,Horava3}.
Here, we do not go into the details of this issue. The detailed
balanced theories show simpler quantum behavior because they have
simpler renormalization properties. However, as is shown in
\cite{Sot}, if one relaxes this condition, the resulting action with
extra allowed terms is well-behavior enough to recover the models
with detailed balance. Another feature of HL theory is its known
inconsistency problems such as its instabilities, ghost scalar modes
and the strong coupling problem. Indeed, by perturbation of this
theory around its IR regime one can show that it suffers from some
instabilities and fine-tunings that may not be removed by usual
tricks such as analytic continuation. Since our study in this
article is done at the background level, such issues are beyond our
discussion. However, a detailed review of this topic can be found in
\cite{Bog}. On the other hand, there are some extensions of the
initial version of the HL gravity theory that deal with such
problems. Some of these are: \cite{Thom}, in which a projectable
$U(1)$ symmetric soft-breaking detailed balance condition model is
considered and it is shown that the resulting theory displays
anisotropic scaling at short distances while almost all features of
GR are recovered at long distances. The non-projectable model
without detailed balance condition is studied in
\cite{non-projectable2}, where it is proved that only
non-projectable model is free from instabilities and strong
coupling. The $U(1)$ symmetric non-projectable version of the HL
gravity is studied in \cite{Zhu}, in which a coupling of the theory
with a scalar field is also considered and it is shown that all the
problems that the original theory suffers from, will be disappeared.
Finally, a progress report around all of the above mentioned issues
has been reviewed in \cite{Rev}.

In this paper we consider a Friedmann-Robertson-Walker (FRW)
cosmological model coupled to a self-interacting scalar field, in
the framework of a projectable HL gravity without detailed balance
condition. The basis of our work to deal with this issue is through
its representation with minisuperspace variables. Minisuperspace
formulation of classical and quantum HL cosmology is studied in some
works, see for instance \cite{vakili&kord,apv} and
\cite{Ber}-\cite{Chris}. Also, quantization of the HL theory without
restriction to a cosmological background is investigated for
instance in \cite{Li}, in which the quantization of two-dimensional
HL theory without the projectability condition is considered, and
\cite{wang}, where a $(1+1)$-dimensional projectable HL gravity is
quantized.

Here, we first construct a suitable form for the HL action and then
will add a self-coupling scalar field to it. For the flat FRW model
and in some special cases, the classical solutions are presented and
their singularities are investigated. We then construct the
corresponding quantum cosmology based on the canonical approach of
Wheeler-DeWitt (WDW) theory to see how things may change their
behavior if the quantum mechanical considerations come into the
model.

\section{The model outline}

To study the FRW cosmology within the framework of HL gravity, let
us start by its geometric structure which in a quasi-spherical polar
coordinate the space time metric is assumed to be

\begin{equation}\label{A}
ds^2=-N^2(t)dt^2+a^2(t)\left[\frac{dr^2}{1-kr^2}+r^2\left(d\vartheta^2+\sin^2\vartheta
d\varphi\right)\right],\end{equation}where $N(t)$ is the lapse
function, $a(t)$ the scale factor and $k=1$, $0$ and $-1$
corresponds to the closed, flat and open universe respectively. In
terms of the ADM variables the above metric takes the form

\begin{equation}\label{B}
g_{\mu \nu}(t,{\bf x})=\left(%
\begin{array}{cc}
-N^2(t) & {\bf 0} \\
{\bf 0} & h_{ab} \\
\end{array}%
\right),
\end{equation}where \[h_{ab}=a^2(t)\mbox{diag}\left(\frac{1}{1-kr^2},r^2,r^2\sin^2\vartheta\right),\]is
the intrinsic metric induced on the spatial $3$-dimensional
hypersurfaces. The gravitational part of the HL action, without the
detailed balance condition, is given by $S_{HL}=S_K+S_V$, where
$S_K$ is its kinetic part

\begin{equation}\label{C}
S_K\sim \int dtd^3\mathbf{x}\,N\sqrt{h}\;\left(K_{ab}K^{ab}-\lambda
K^2\right),
\end{equation}where $h$ is the determinant of $h_{ab}$ and $\lambda$ is a correction constant to the usual
GR due to HL theory. Also, $K_{ab}$ is the extrinsic curvature (with
trace $K$) defined as
\[K_{ab}=\frac{1}{2N}\left(N_{a|b}+N_{b|a}-\frac{\partial
h_{ab}}{\partial t}\right),\] where $N_{a|b}$ denotes the covariant
derivative with respect to $h_{ab}$. Since for the FRW metric all
components of the shift vector are zero, a simple calculation based
on the above definition results in
$K_{ab}K^{ab}=\frac{3\dot{a}^2}{N^2 a^2}$ and
$K=-\frac{3\dot{a}}{Na}$, where a dot represents differentiation
with respect to $t$. Going back to the action, its potential part is
in the form

\begin{equation}\label{D}
S_V=-\int dtd^3\mathbf{x} N \sqrt{h} V[h_{ij}].
\end{equation}According to the relation (\ref{In1}) and because of the anisotropic
scaling of space and time coordinates, their dimensions are
different as $[\mathbf{x}]=[\kappa]^{-1}$ and $[t]=[\kappa]^{-z}$,
where the $[\kappa]$ is a symbol of dimension of momentum. In this
sense, the dimension of the metric, lapse function and shift vector
will be $[\gamma_{ij}]=[N]=1$ and $[N^i]=[\kappa]^{z-1}$. Therefore,
the potential term in a three-dimensional space has the dimension
$\left[V[h_{ij}]\right]=[\kappa]^{z+3}$. So, according to such a
dimensional analysis one may argue that for special case $z=3$, the
potential $V[h_{ij}]$ consists of the following terms of the Ricci
tensor, Ricci scalar and their covariant derivatives of dimension
$[\kappa]^6$, \cite{Visser}

\begin{eqnarray}\label{E}
V[h_{ij}]&=&g_0\zeta^6+g_1\zeta^4R+g_2\zeta^2R^2+g_3\zeta^2R_{ij}R^{ij}\nonumber\\
&&+g_4R^3+g_5RR_{ij}R^{ij}+g_6R_{ij}R^{jk}R_k^{\;i}\nonumber\\
&&+g_7R\nabla^2R+g_8\nabla_iR_{jk}\nabla^iR^{jk},
\end{eqnarray}where $g_{i}$ ($i=0,...,8$) are dimensionless coupling constants
come from HL correction to usual GR and $ \zeta$ with dimension
$[\zeta]=[\kappa]$ is introduced to make the constants $g_i$
dimensionless. Under these conditions, the gravitational part of the
HL theory that we shall consider hereafter has the following form,
\cite{Sot}, \cite{Visser}, \cite{Pitelli&Saa} and \cite{Saridakis}

\begin{eqnarray}\label{F}
S_{HL}&=&\frac{M_{PL}^2}{2}\int_\mathcal{M}dtd^3\mathbf{x}\;N\sqrt{h}\Big[K_{ij}K^{ij}-\lambda K^2+R-2\Lambda\nonumber\\
&&\qquad-\frac{g_2}{M_{PL}^2}R^2-\frac{g_3}{M_{PL}^2}R_{ij}R^{ij}-\frac{g_4}{M_{PL}^4}R^3
-\frac{g_5}{M_{PL}^4}RR_{ij}R^{ij}\nonumber\\
&&\qquad-\frac{g_6}{M_{PL}^4}R_{ij}R^{jk}R^i_{\;k}
-\frac{g_7}{M_{PL}^4}R\nabla^2R-\frac{g_8}{M_{PL}^4}\nabla_iR_{jk}\nabla^iR^{jk}\Big],
\end{eqnarray}
in which $M_{PL}=\frac{1}{\sqrt{8\pi G}}$ and we have set $c=1$,
$\zeta=1$, $\Lambda=g_0M_{Pl}^2/2$ and $g_1=-1$.

Now, let us consider a scalar field minimally coupled to gravity but
has a non-linear interaction with itself by a coupling function
$F(\phi)$ \cite{vakili}. The action of such a scalar field may be
written as

\begin{equation}\label{G}
S_{\phi}=\int_\mathcal{M}
d^4\mathbf{x}\sqrt{-g}F(\phi)g^{\mu\nu}\partial_{\mu}\phi\partial_{\nu}\phi.
\end{equation}The existence of a scalar field in a gravitational theory can address many issues in
cosmology such as spatially flat and accelerated expanding universe
at the present time, inflation, dark matter and dark energy. The
action of the scalar field considered here has the same form as that
in usual cosmological models with general covariance. However, in
the HL gravity the Lorentz symmetry is broken in such a way that
various higher spatial derivatives will appear in the gravitational
part of the action. Therefore, one expects this feature to be
considered when constructing the action of the scalar field. This
means that we may be able to add higher spatial derivatives of the
scalar field into its action. One of the possible types of such
actions for scalar field is presented in \cite{Eli} as

\begin{equation}\label{G1}
S_{\phi}=\int d^4\mathbf{x}\sqrt{-g}
N\left[\frac{1}{N^2}\left(\dot{\phi}-N^i\partial_i
\phi\right)^2-{\cal V}(\partial_i \phi,\phi)\right],
\end{equation}where the the potential function ${\cal V}$ can in general contain arbitrary combinations of $\phi$
and its spatial derivatives. However, as emphasized in \cite{Eli},
in homogeneous and isotropic cosmological settings like FRW metric,
we have $N^i=0$ and the cosmological ansatz for the scalar field is
$\phi=\phi(t)$. In such a cases since $\partial_i \phi=0$, the
function ${\cal V}$ in the scalar action reduces effectively to a
usual potential that vanishes for a free scalar field. In this
respect, the action (\ref{G}), we presented for a self-interacting
scalar field in the theory appears to be based on physical grounds.

The total action may now be written by adding the HL and scalar
field actions as $S=S_{HL}+S_{\phi}=\int dt
L[a,\phi,\dot{a},\dot{\phi}]$. Having at hand the actions (\ref{F})
and (\ref{G}), by substituting the metric (\ref{A}) into them, we
are led to the following effective Lagrangian in terms of the
minisuperspace variables $(a(t), \phi(t))$ \footnote{The Planck mass
can be absorbed in time term and so we may re-scale the time as
$t\rightarrow \frac{t}{t_{PL}}$, where $t_{PL}=\frac{1}{M_{PL}}$ is
the Planck time. In this sense, in what follows, by $t$, we mean the
dimensionless quantity $\frac{t}{t_{PL}}$. Specially, the figures
are plotted in terms of this quantity.}:

\begin{equation}\label{H}
L[a,\phi,\dot{a},\dot{\phi}]=N\left(-\frac{a\dot{a}^2}{N^2}+g_ca-g_{\Lambda}a^3-\frac{g_r}{a}-\frac{g_s}{a^3}
+\frac{1}{N^2}F(\phi)a^3\dot{\phi}^2\right),
\end{equation}in which the new coefficients are defied as \cite{Kei}

\begin{equation}\label{I}
g_c=\frac{6k}{3(3\lambda-1)},\hspace{0.3cm}g_{\Lambda}=\frac{2\Lambda}{3(3\lambda-1)},\hspace{0.3cm}g_r=\frac{12k^2(3g_2+g_3)}{3(3\lambda-1)M^2_{PL}}
,\hspace{0.3cm}
g_s=\frac{24k^3(9g_4+3g_5+g_6)}{3(3\lambda-1)M^4_{PL}}.\end{equation}
The momenta conjugate to each of the dynamical variables can be
obtained by definition $p_q=\frac{\partial L}{\partial \dot{q}}$
with results $p_{a}=-\frac{2a\dot{a}}{N}$ and
$p_{\phi}=\frac{2}{N}F(\phi)a^3\dot{\phi}$. In terms of the these
momenta, the total Hamiltonian reads

\begin{eqnarray}\label{J}
H=N\mathcal{H}=N\left(-\frac{p_a^2}{4a}+\frac{p_{\phi}^2}{4a^3F(\phi)}-g_ca+g_{\Lambda}a^3+\frac{g_r}{a}+\frac{g_s}{a^3}\right).
\end{eqnarray}
As it should be, the lapse function appears as a Lagrange multiplier
in the Hamiltonian which means that the Hamiltonian equation for
this variable yields the constraint equation ${\cal H}=0$. At
classical level this constraint is equivalent to the Friedmann
equation. As we shall see later, this constraint also plays a key
role in forming the basic equation of quantum cosmology, that is,
the WDW equation.

\section{Classical cosmology}
In this section we intend to study the classical cosmological
solutions of the HL model whose Hamiltonian is given by equation
(\ref{J}). In Hamiltonian approach, the classical dynamics of each
variable is determined by the Hamilton equation $\dot{q}=\left\{q,
H\right\}$, where $\left\{.,.\right\}$ is the Poisson bracket. So,
we get

\begin{eqnarray}\label{K}
\left\{
\begin{array}{cl}
\dot{a}&=-\frac{Np_a}{2a},\\
\dot{p}_a&=N\left(-\frac{p_a^2}{4a^2}+g_c-3g_{\Lambda}a^2+\frac{g_r}{a^2}+\frac{3g_s}{a^4}
+\frac{3p_{\phi}^2}{4a^4F(\phi)}\right),\\
\dot{\phi}&=\frac{Np_{\phi}}{2a^3F(\phi)},\\
\dot{p}_{\phi}&=\frac{Np_{\phi}^2}{4a^3}\frac{F'}{F^2},
\end{array}
\right.
\end{eqnarray}where $F'=\frac{dF(\phi)}{d\phi}$. Before attempting to solve the above system of
equations, we must choose a time gauge by fixing the Lapse function.
Without this, we will face with the problem of under-determinacy
which means that there are fewer equations than unknowns. So, let us
fix the lapse function as $N=a^n(t)$, where $n$ is a constant. With
this time gauge, by eliminating $p_{\phi}$ from two last equations
of (\ref{K}), we obtain the following equation for $\phi$:

\begin{equation}\label{L}
\frac{\ddot{\phi}}{\dot{\phi}}+\frac{1}{2}\frac{F'(\phi)}{F(\phi)}\dot{\phi}+(3-n)\frac{\dot{a}}{a}=0,
\end{equation}which seems to be a conservation law for the scalar
field. This equation can easily be integrated with result

\begin{equation}\label{M}
\dot{\phi}^2F(\phi)=Ca^{2(n-3)},
\end{equation}where $C$ is an integration constant. Now, to obtain a differential equation for the scale factor, let us
eliminate the momenta from the system (\ref{K}) from which and also
using (\ref{M}) we will arrive at

\begin{equation}\label{N}
\dot{a}^2+a^{2n}\left(g_c-g_{\Lambda}a^2-\frac{g_r}{a^2}-\frac{g_s+C}{a^4}\right)=0,
\end{equation}which is equivalent to the Hamiltonian constraint
$H=0$. The last differential equation we want to derive from the
above relations is an equation between $a$ and $\phi$ whose
solutions give us the classical trajectories in the plane $a-\phi$.
This may be done by removing the time parameter between (\ref{M})
and (\ref{N}) which yields

\begin{equation}\label{O}
F(\phi)\left(\frac{d\phi}{da}\right)^2=C
a^{-6}\left(-g_c+g_{\Lambda}a^2+\frac{g_r}{a^2}+\frac{g_s+C}{a^4}\right)^{-1}.
\end{equation}The non-dependence of this equation on the parameter $n$ indicates that although
different time gauges lead to different functions for scale factor
and scalar field, the classical trajectories are independent of
these gauges. On the other hand, from now on, to make the model
simple and solvable, we take a polynomial coupling function for the
scalar field as $F(\phi)=\lambda \phi^m$.

In general, the above equations do not seem to have analytical
solutions, so in what follows we restrict ourselves to some special
cases for which we can obtain analytical closed form solutions for
the above field equations.

\subsection{Flat universe with cosmological constant: $k=0$, $\Lambda \neq 0$}
\begin{figure}
\centering
\includegraphics[width=0.45\textwidth]{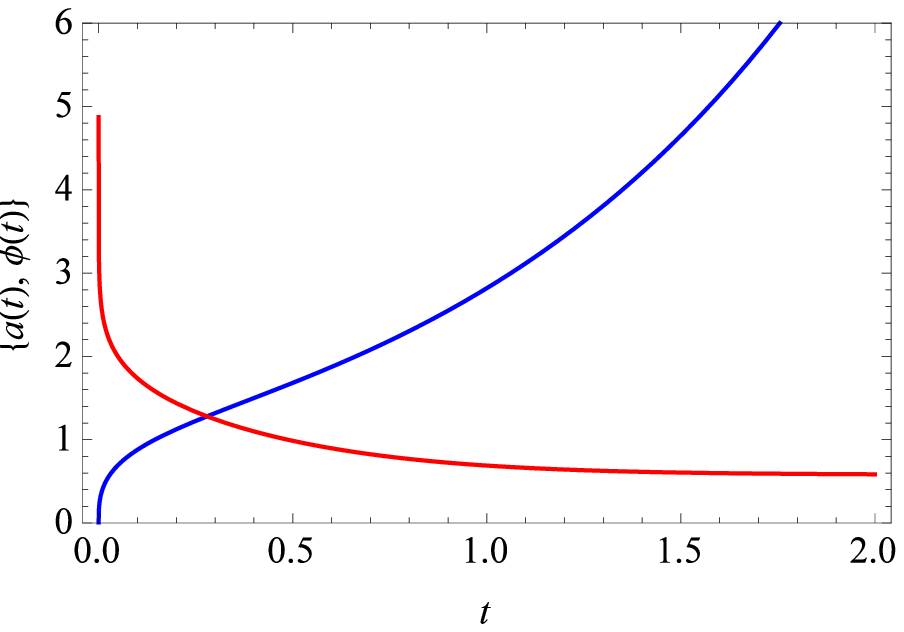}
\includegraphics[width=0.45\textwidth]{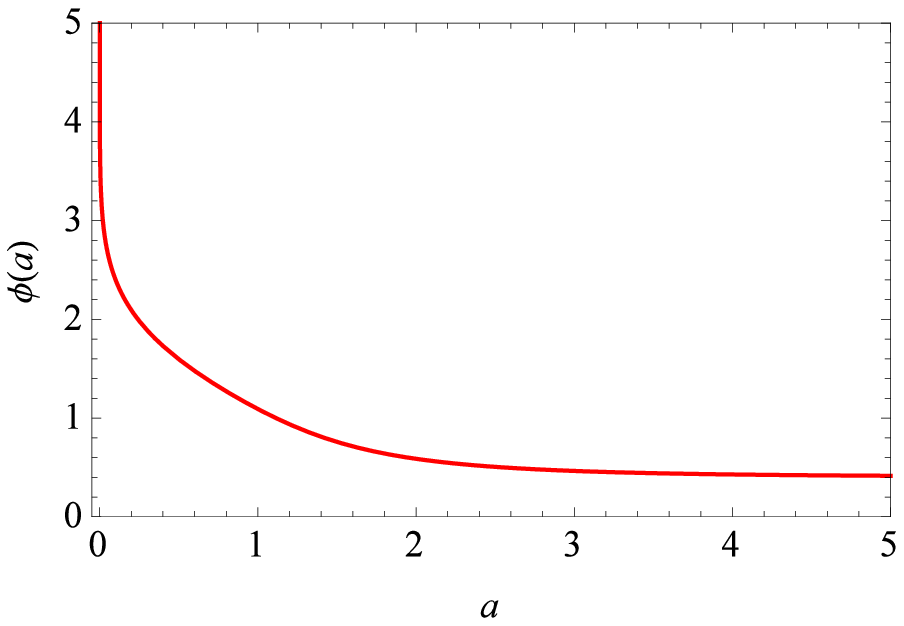}\\
\caption{Left: the qualitative dynamical behavior of the scale
factor (blue line) and scalar field (red line) in flat universe.
Right: classical trajectory in the plane $a-\phi$. The figures are
plotted for numerical values: $g_{\Lambda}=1$, $C=5$, $\lambda=1$,
$m=2$ and $\phi_c=\frac{2}{5}$.} \label{fig1}
\end{figure}

For a flat universe the coefficients $g_c$, $g_r$ and $g_s$ vanish.
So, if we choose the time parameter corresponding to the gauge
$N=1$, or equivalently $n=0$, the Friedmann equation (\ref{N}) reads

\begin{equation}\label{P}
\dot{a}^2=g_{\Lambda}a^2+\frac{C}{a^4},
\end{equation}with solution

\begin{equation}\label{Q}
a(t)=\left(\frac{C}{g_{\Lambda}}\right)^{\frac{1}{6}}
\sinh^{\frac{1}{3}}(3\sqrt{g_\Lambda}t),
\end{equation}where the integration constant is adopted in such a way that the singularity occurs
at $t=0$, which means that $a(t=0)=0$. Now, let us to find an
expression for the scalar field. As we mentioned before, we consider
its self-coupling function in the form of $F(\phi)=\lambda \phi^m$.
With this choice for the function $F(\phi)$, equation (\ref{M})
takes the form

\begin{equation}\label{R}
\phi^{\frac{m}{2}}d\phi=\pm\sqrt{\frac{C}{\lambda}}\frac{dt}{a^3(t)},
\end{equation}which with the help of equation (\ref{Q}), we are able
to integrate it and find the time evolution of the scalar field as

\begin{equation}\label{S}
\phi(t)=\left[\phi_c^{\frac{m+2}{2}}-\frac{m+2}{6\sqrt{\lambda}}\;
\ln\left(\tanh\frac{3\sqrt{g_\Lambda}\;t}{2}\right)\right]^{\frac{2}{m+2}},\qquad
m\neq-2,
\end{equation}where the integration constant $\phi_c$ is chosen such
that $\lim_{t\rightarrow\infty}\phi(t)=\phi_c$. In figure
\ref{fig1}, we have plotted the behavior of the scale factor and the
scalar field. As this figure shows, the universe begins its
evolution with a big-bang singularity (zero size for $a(t)$) at
$t=0$ where the scalar field blows up there. As time goes on, while
the universe expands (with positive acceleration) until it finally enters to a de Sitter phase,
that is, we have $a(t)\sim e^{\sqrt{g_{\Lambda}}t}$, as
$t\rightarrow+\infty$, the scalar field eventually tends to a
constant value. We can also follow this behavior by studying the
classical trajectories in the plane $a-\phi$. To do this, we need to
extract the scalar field in terms of the scale factor from equation
(\ref{O}) which now can be written as

\begin{equation}\label{T}
\phi^m\left(\frac{d\phi}{da}\right)^2=\frac{C}{\lambda}\frac{1}{a^2(g_{\Lambda}a^6+C)},
\end{equation}whose integration yields

\begin{equation}\label{U}
\phi(a)=\left[\phi_c^{\frac{m+2}{2}}+\frac{m+2}{6\sqrt{\lambda}}\;
\ln\frac{\sqrt{g_{\Lambda}a^6+C}-\sqrt{C}}{\sqrt{g_{\Lambda}}\;a^3}\right]^{\frac{2}{m+2}},\qquad
m\neq-2.
\end{equation}In figure \ref{fig1}, we also plotted the above
expression for typical numerical values of the parameters. How the
scale factor varies with respect to the scalar field, or vice versa,
can also be seen from this figure. We will return to this classical
trajectory again when looking at the quantum model to investigate
whether the peaks of the wave function correspond to these paths.

\subsection{Non-flat universe with zero cosmological constant: $k\neq 0$, $\Lambda =0$}

\begin{figure}
\centering
\includegraphics[width=0.45\textwidth]{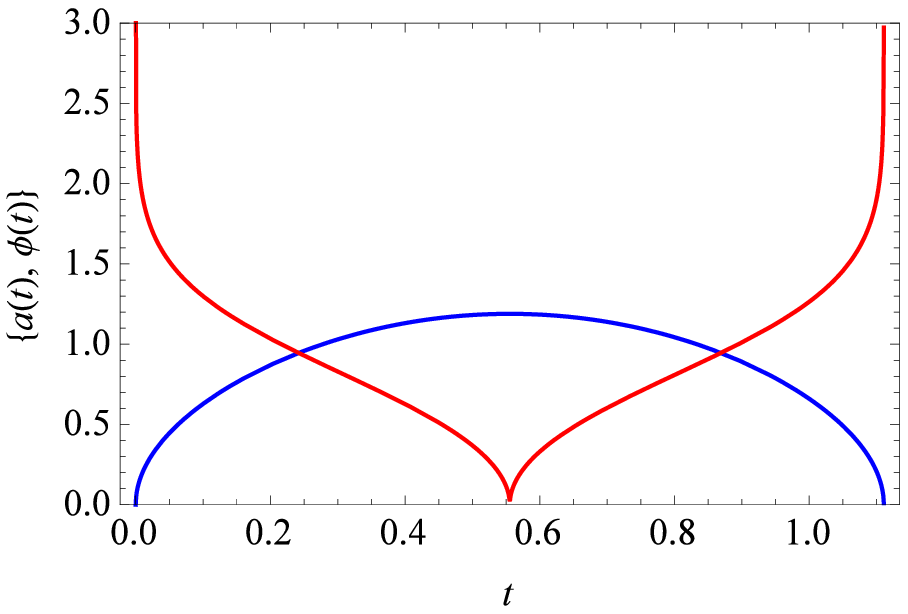}
\includegraphics[width=0.45\textwidth]{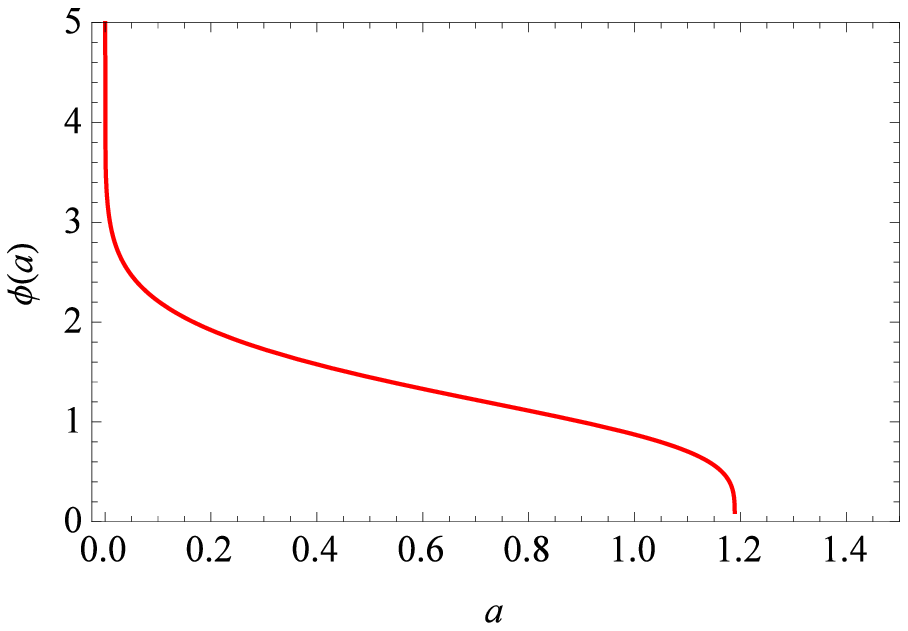}\\
\caption{Left: the qualitative behavior of the scale factor (blue line) and scalar field (red line) when
$\Lambda=0$ and $g_r=0$. Right: the classical trajectory
in the plane $a-\phi$. The figures are plotted for the numerical values: $g_c=2$, $g_s=1$, $C=3$, $\lambda=1$, $m=2$ and $\phi_c=0$.}
\label{fig2}
\end{figure}

In this subsection, we consider another special case in which while
the curvature index is non-zero but the cosmological constant is
equal to zero. This means that $g_c,g_s,g_r \neq 0$ and
$g_{\Lambda}=0$. Under these conditions if we take an evolution
parameter corresponding to the lapse function $N(t)=a(t)$, (or
$n=1$), the Friedmann equation (\ref{N}) will be

\begin{equation}\label{W}
\dot{a}^2+g_ca^2-g_r-\frac{g_s+C}{a^2}=0.
\end{equation}Before trying to solve this equation, we should note a
point about the selected lapse function. Unlike the case in the
previous subsection in which our time parameter with $N=1$, was
indeed the usual cosmic time, here with $N=a(t)$, $t$ is just a
evolution or clock parameter in terms of which the evolution of all
dynamical variables is measured. However, one may translate the
final results in terms of the cosmic time $\tau$, using its relation
with the time parameter $t$, that is, $d\tau=N(t)dt$.

the general solution to the equation (\ref{W}) is

\begin{equation}\label{X}
a(t)=\sqrt{\alpha(1-\cos\omega t)+\beta\sin\omega t},
\end{equation}where $\alpha=\frac{g_r}{2g_c}$, $\beta=\sqrt{\frac{g_s+C}{g_c}}$
and $\omega=2\sqrt{g_c}$. To express this and the following
relations in a simpler form let us take $g_r=0$, which is equivalent
to $g_3=-3g_2$ in (\ref{I}). Also, we assume that $\lambda > 1/3$
and $9g_4+3g_5+g_6 >0$, so that $\mbox{sign}(g_c,
g_s)=\mbox{sign}(k)$. In the following, we will present the
solutions for the closed universe $k=+1$. The open ($k=-1$)
counterpart of the solutions can be obtained via making small
changes by replacing the trigonometric functions with their
hyperbolic counterparts. Therefore, by applying, again, the initial
condition $a(t=0)=0$, we have

\begin{equation}\label{Y}
a(t)=\left(\frac{g_s+C}{g_c}\right)^{\frac{1}{4}}\sqrt{\sin(2\sqrt{g_c}t)}.
\end{equation}The time dependence of the scalar field can also be deduced from equation (\ref{M}) which for the present case has the solution

\begin{equation}\label{Z}
\phi(t)=\left[\phi_c^{\frac{m+2}{2}}+\frac{m+2}{4}\sqrt{\frac{C}{(g_s+C)\lambda}}\ln\tan(\sqrt{g_c}\,t)\right]^{\frac{2}{m+2}},\qquad m\neq-2.
\end{equation}Finally, what remains is the classical trajectories in
the plane $a-\phi$ which as before may be obtained from (\ref{O}) with result

\begin{equation}\label{AB}
\phi(a)=\left[\frac{m+2}{4}\sqrt{\frac{C}{(g_s+C)\lambda}}\;\ln\frac{\sqrt{g_s+C}+\sqrt{g_s+C-g_ca^4}}{\sqrt{g_c}\;a^2}\right]^{\frac{2}{m+2}}.
\end{equation}In figure \ref{fig2}, we have shown the time behavior of
the scale factor and the scalar field. As is clear from this figure,
the universe begins its decelerated expansion from a singularity
where the size of the universe is zero and the value of the scalar
field is infinite. As the scale factor expands to its maximum value,
the scalar field decreases to zero. Then, by re-collapsing the scale
factor to a big-crunch singularity, the scalar field again increases
until it blows where the scale factor vanishes. The behavior of the
scale factor and the scalar field in the plane $a-\phi$ is also
plotted in this figure. This figure also clearly shows the divergent
behavior of the scalar field where the scale factor is singular.

\subsection{Early universe}
\begin{figure}
\centering
\includegraphics[width=0.45\textwidth]{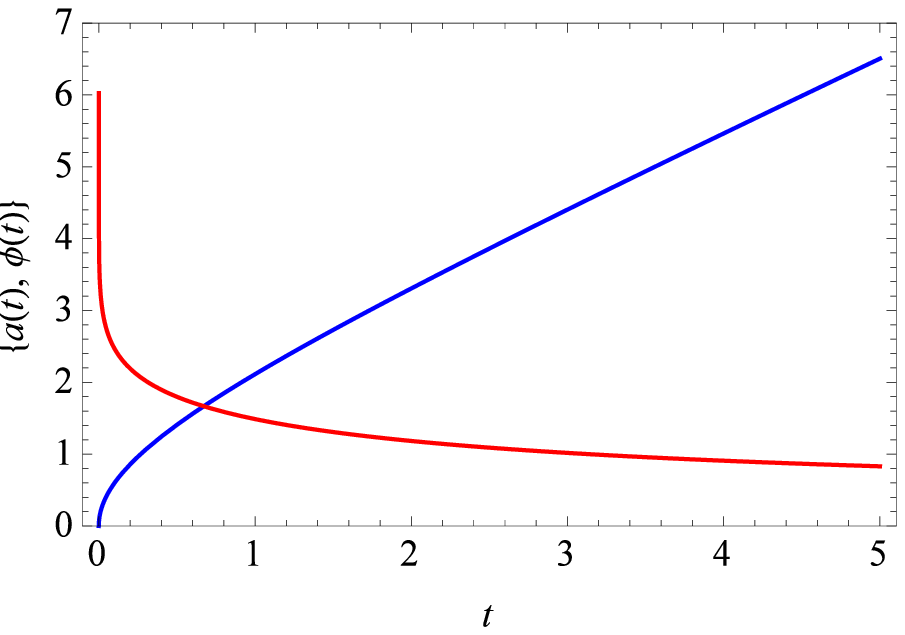}
\includegraphics[width=0.45\textwidth]{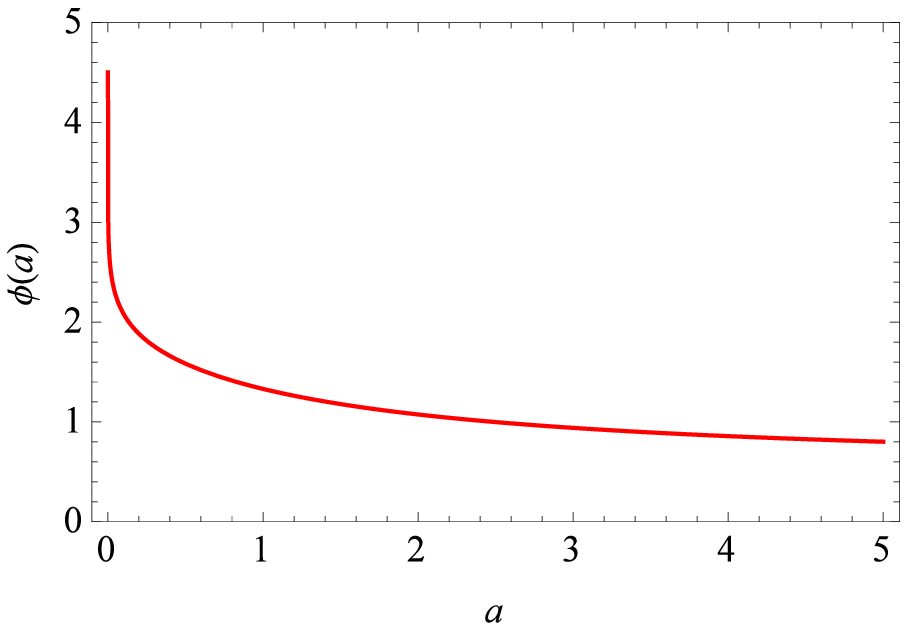}\\
\caption{Left: the behavior of the scale factor (blue line) and scalar field
(red line) in the early times of cosmic evolution. Right: the corresponding classical trajectory
in the plane $a-\phi$. The figures are plotted for numerical values:
$g_r=1$, $g_s=2$, $C=1$, $\lambda=1$, $m=2$ and $\phi_c=\frac{1}{2}$.}
\label{fig3}
\end{figure}

In this subsection we consider the dynamics of the universe in the
early times of cosmic evolution when the scale factor is very small.
For a such a situation the Friedmann equation ({\ref{N}}) (again in
the gauge $N=a(t)$) takes the form

\begin{equation}\label{AC}
\dot{a}^2=g_r+\frac{g_s+C}{a^2},
\end{equation}in which we omit the terms containing $a^2$ and $a^4$.
It is easy to derive the scale factor from this equation as

\begin{equation}\label{AD}
a(t)=\left[g_r\;(t+\delta)^2-\frac{g_s+C}{g_r}\right]^{\frac{1}{2}},
\end{equation}where $\delta=\frac{\sqrt{g_s+C}}{g_r}$. This equation shows
that, regardless of whether the curvature index is positive or negative,
the universe has a power law expansion in the early times of its evolution coming from a big-bang singularity.
Following the same steps we took in the previous sections will lead us to the following expressions for the scalar field and the
classical trajectory

\begin{equation}\label{AE}
\phi(t)=\left[\phi_c^{\frac{m+2}{2}}-\frac{m+2}{4}\sqrt{\frac{C}{(g_s+C)\lambda}}\;
\ln\frac{g_r(t+\delta)-\sqrt{g_s+C}}{g_r(t+\delta)+\sqrt{g_s+C}}\right]^{\frac{2}{m+2}},\qquad m\neq-2,\end{equation}and

\begin{equation}\label{AF}
\phi(a)=\left[\phi_c^{\frac{m+2}{2}}+\frac{m+2}{2}\sqrt{\frac{C}{(g_s+C)\lambda}}\;
\ln\frac{\sqrt{g_ra^2+g_s+C}+\sqrt{g_s+C}}{\sqrt{g_r}\;a}\right]^{\frac{2}{m+2}}.\end{equation}
As before we summarized all the results of this subsection in figure \ref{fig3}, which illustrates the evolution and singularity of the
dynamical variables.

\subsection{Late time expansion}
Another important issue in cosmological dynamics is the late time behavior of the universe. In this limit the Friedmann
equation (\ref{N}), in the gauge $N=1$, has the form

\begin{equation}\label{AG}
\dot{a}^2+g_c-g_{\Lambda}a^2=0,
\end{equation}where here we have neglected the terms $1/a^2$ and
$1/a^4$. It is easy to see that this equation is solved by the following functions

\begin{equation}\label{AH}
a(t)=\frac{1}{2g_{\Lambda}}\left(e^{\sqrt{g_{\Lambda}}t}+g_c
g_{\Lambda}e^{-\sqrt{g_{\Lambda}}t}\right),\hspace{5mm}
a(t)=\frac{1}{2g_{\Lambda}}\left(g_c g_{\Lambda}e^{\sqrt{g_{\Lambda}}t}+e^{-\sqrt{g_{\Lambda}}t}\right),
\end{equation}both of which enter the de Sitter phase

\begin{equation}\label{AI}
a(t)\sim e^{\sqrt{g_{\Lambda}}t},
\end{equation}when $t\rightarrow +\infty$. Similar to the calculations of the preceding
sections, we can obtain the following expression for the scalar field

\begin{equation}\label{AJ}
\phi(t)=\left[\phi_c^{\frac{m+2}{2}}+\frac{(m+2)g_{\Lambda}}{4g_c}\sqrt{\frac{C}{\lambda\,g_c}}\,
\left(\frac{\tanh(\sqrt{g_\Lambda}\;t)}{\cosh(\sqrt{g_\Lambda}\;t)}
+2\arctan\left(e^{\sqrt{g_\Lambda}\;t}\right)\right)\right]^{\frac{2}{m+2}},
\end{equation}which tends to a constant value as $t\rightarrow
+\infty$.

\section{Canonical quantization of the model}
As we mentioned before, HL gravity is a generalization of the usual
GR at UV regimes in such a way that in the low energy limits the
standard GR is recovered. Therefore, from a cosmological point of
view, one may obtained some nonsingular bouncing solutions. From
this perspective, this theory may be considered as an alternative to
inflation since it is expected it might solve the flatness and
horizon problems and generate scale invariant perturbations for the
early universe without the need of exponential expansion usually
used in the inflationary theories \cite{Rabert}.

On the other hand, at the background (non-perturbation) level almost
all solutions to the Einstein field equations exhibit different
kinds of singularities. On this basis, cosmological solutions along
with conventional matter fields are no exception to this rule and
mainly exhibit big bang type singularities. Any hope to eliminate
these singularities would be in the development of a concomitant and
conducive quantum theory of gravity. In the absence of a complete
theory of quantum gravity, it would be useful to describe the
quantum state of the universe in the context of quantum cosmology,
in which based on the canonical quantization procedure, the
evolution of the universe is described by a wave function in the
minisuperspace. In other words, in this view, the system in question
will be reduced to a conventional quantum mechanical system. In what
follows, according to this procedure, we are going to overcome the
singularities that appear in the classical model.

Now let us focus our attention on the quantization of the model
described in the previous section. To do this, we start with the
Hamiltonian (\ref{J}). As we know, the lapse function in such
Hamiltonians appears itself as a Lagrange multiplier, so we have the
Hamiltonian constraint $H=0$. This means that application of the
canonical quantization procedure demands that the quantum states of
the system (here the universe) should be annihilated by the quantum
(operator) version of $H$, which yields the WDW equation
$\hat{H}\Psi(a,\phi)=0$, where $\Psi(a,\phi)$ is the wave function
of the universe. To obtain the differential form of this equation,
if we use the usual representation $P_q\rightarrow
\partial_q$, we are led to
the following WDW equation

\begin{eqnarray}\label{AK}
&&\frac{1}{4a}\left(\frac{\partial^2}{\partial a^2}
+\frac{\beta}{a}\frac{\partial}{\partial a}\right)\Psi(a,\phi)
-\frac{1}{4a^3F(\phi)}\left(\frac{\partial^2}{\partial\phi^2}
+\frac{\kappa
F'(\phi)}{F(\phi)}\frac{\partial}{\partial\phi}\right)\Psi(a,\phi)\nonumber\\\label{WD}
&&+\left(-g_ca+g_{\Lambda}a^3+\frac{g_r}{a}+\frac{g_s}{a^3}\right)\Psi(a,\phi)=0,
\end{eqnarray}where the parameters $\beta$ and $\kappa$ represent the ambiguity in the ordering of
factors $(a,P_a)$ and $(\phi, P_{\phi})$ respectively. It is clear
that there are lots of the possibilities to choose this parameters.
For example with $\beta=\kappa=0$, we have no factor ordering, with
$\beta=\kappa=1$, the kinetic term of the Hamiltonian takes the form
of the Laplacian $-\frac{1}{2}\nabla^2$, of the minisuperspace. In
general, as is clear from the WDW equation, the resulting form of
the wave function depends on the chosen factor ordering \cite{Ste}.
However, it can be shown that the factor-ordering parameter will not
affect semiclassical calculations in quantum cosmology \cite{Peg},
and so for convenience one usually chooses a special value for it in
the special models.

As the first step in solving the equation (\ref{AK}), let us
separate the variables into the form
$\Psi(a,\phi)=\psi(a)\Phi(\phi)$, which yields the following
differential equations for the functions $\psi(a)$ and $\Phi(\phi)$:

\begin{equation}\label{AL}
\frac{d^2\psi(a)}{da^2}+\frac{\beta}{a}\frac{d\psi(a)}{da}
+4\left(-g_ca^2+g_{\Lambda}a^4+g_r+\frac{g_s+w^2}{a^2}\right)\psi(a)=0,
\end{equation}and

\begin{equation}\label{AM}
\frac{d^2\Phi(\phi)}{d\phi^2}+\frac{\kappa
F'(\phi)}{F(\phi)}\frac{d\Phi(\phi)}{d\phi}+4w^2F(\phi)\Phi(\phi)=0,
\end{equation}with $w$ being a separation constant. As in the
classical case, here we examine the analytical solutions of the
above equations in a few specific cases. Moreover, we assume that
the wave functions are supposed to obey the boundary conditions

\begin{equation}\label{AN}
\Psi(a=0,\phi)=0,\hspace{5mm} \mbox{Dirichlet B.C.},
\end{equation}

\begin{equation}\label{AO}
\qquad\left.\frac{\partial\Psi(a,\phi)}{\partial
a}\right|_{a=0}=0,\hspace{5mm} \mbox{Neumann B.C.},
\end{equation}where the first condition is called the Dewitt boundary
condition to avoid the singularity in the quantum domain. In what
follows we will deal with the resulting quantum cosmology in the
same special cases that we have already examined the classical
solutions.

\subsection {Flat quantum universe with cosmological constant: $k=0$,
$\Lambda\neq 0$}

In this case by selecting $\beta=-2$, the equation (\ref{AL}) reads
as

\begin{equation}\label{AP}
\frac{d^2\psi(a)}{da^2}-\frac{2}{a}\frac{d\psi(a)}{da}+4\left(g_{\Lambda}a^4+\frac{w^2}{a^2}\right)\psi(a)=0,
\end{equation}the solutions of which in terms of Bessel functions
$J_{\nu}(z)$ and $Y_{\nu}(z)$ are as follows

\begin{equation}\label{AQ}
\psi(a)=C_1\;a^{\frac{3}{2}}\;\mathrm{J}_{\frac{i}{6}\sqrt{16w^2-9}}\left(\frac{2}{3}\sqrt{g_{\Lambda}}\;a^3\right)
+C_2\;a^{\frac{3}{2}}\;\mathrm{Y}_{\frac{i}{6}\sqrt{16w^2-9}}\left(\frac{2}{3}\sqrt{g_{\Lambda}}\;a^3\right),
\end{equation}where $C_1$ and $C_2$ are the integration constants. In the case where the order of Bessel functions is
imaginary, ($w^2>9/16$), both of them satisfy the DeWitt boundary
conditions and so both integral constants can be non-zero which we
take them here as $C_1=1$ and $C_2=i$.

On the other hand putting the coupling function
$F(\phi)=\lambda\phi^m$, and setting the ordering parameter as
$\kappa=0$, equation (\ref{AM}) takes the form

\begin{equation}\label{AR}
\frac{d^2\Phi(\phi)}{d\phi^2}+4w^2\lambda\phi^m\Phi(\phi)=0,
\end{equation}with solutions

\begin{equation}\label{AS}
\Phi(\phi)=C_3\;\sqrt{\phi}\,\mathrm{J}_{\frac{1}{m+2}}
\left(\frac{4\sqrt{\lambda}w}{m+2}\phi^{\frac{m+2}{2}}\right)
+C_4\;\sqrt{\phi}\,\mathrm{Y}_{\frac{1}{m+2}}
\left(\frac{4\sqrt{\lambda}w}{m+2}\phi^{\frac{m+2}{2}}\right).
\end{equation} Thus, the eigenfunctions of the WDW equation can be written as

\begin{eqnarray}\label{AT}
\Psi_w(a,\phi)&=&\psi_w(a)\Phi_w(\phi)\nonumber\\
&=&a^{\frac{3}{2}}\sqrt{\phi}\;\mathrm{H}^{(1)}_{\frac{i}{6}\sqrt{16w^2-9}}\left(\frac{2}{3}\sqrt{g_{\Lambda}}\;a^3\right)
\,\mathrm{J}_{\frac{1}{m+2}}\left(\frac{4\sqrt{\lambda}w}{m+2}\phi^{\frac{m+2}{2}}\right),
\end{eqnarray}where we choose $C_4=0$, for having well-defined functions in all ranges of
variable $\phi$, and
$\mathrm{H}^{(1)}_{\nu}(z)=\mathrm{J}_{\nu}(z)+i\mathrm{Y}_{\nu}(z)$
are the Hankel functions. We may now write the general solutions to
the DWD equation as a superposition of the above eigenfunctions,
that is

\begin{eqnarray}\label{AU}
\Psi(a,\phi)&=&\int_{D}dw\;f(w)\;\Psi_w(a,\phi)\nonumber\\
&=&a^{\frac{3}{2}}\sqrt{\phi}\int_{D}dw\;f(w)\;
\mathrm{H}^{(1)}_{\frac{i}{6}\sqrt{16w^2-9}}\left(\frac{2}{3}\sqrt{g_{\Lambda}}\;a^3\right)
\,\mathrm{J}_{\frac{1}{m+2}}\left(\frac{4\sqrt{\lambda}w}{m+2}\phi^{\frac{m+2}{2}}\right),
\end{eqnarray}where $f(w)$ is a suitable weight function to construct the quantum
wave packets and $D=(-\infty,-3/4]\cup[+3/4,+\infty)$ is the domain
on which the integral is taken. It is seen that this expression is
too complicated for extracting an analytical closed form for the
wave function and the choice of a function $f(w)$ that leads to an
analytical solution for the wave function is not an easy task.
However, such weight functions in quantum systems can be chosen as a
shifted Gaussian weight function

\begin{equation}\label{AU1}
f(w)=w^p e^{-\sigma(w-w_0)^2},
\end{equation}
which are widely used in quantum mechanics as a way to construct the
localized states. This is because these types of weight factors are
centered about a special value of their argument and they fall off
rapidly away from that center. Due to this behavior the
corresponding wave packet resulting from (\ref{AU}) after
integration, has also a Gaussian-like behavior, i.e. is localized
about some special values of its arguments.

\begin{figure}

\includegraphics[width=0.45\textwidth]{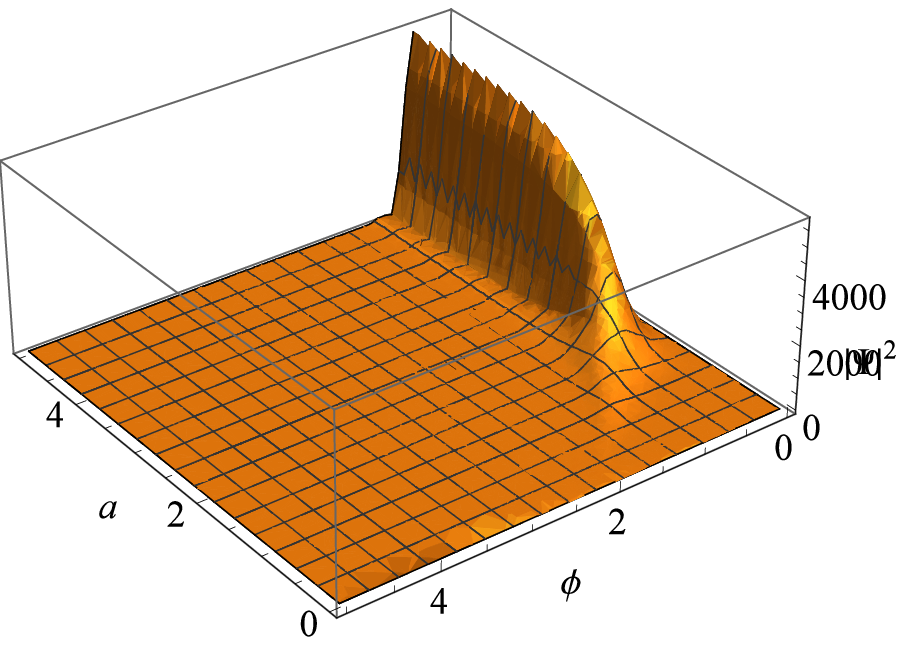}
\includegraphics[width=0.45\textwidth]{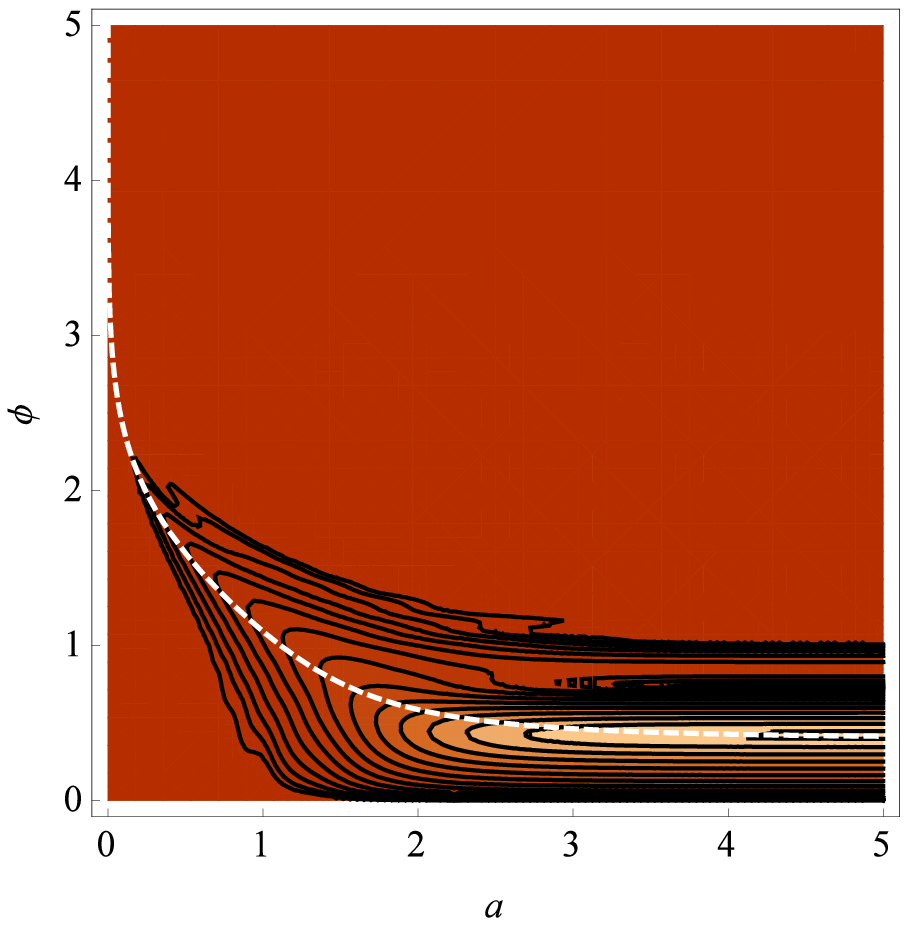}\\
\caption{The figures show the square of the wave function (left) and
its corresponding contour plot (right). Also, dashed line denotes
the classical trajectory of the system in phase space. The figures
are plotted for the numerical values: $g_{\Lambda}=1$, $\lambda=1$,
$m=2$, $p=0$, $\sigma=\frac{1}{10}$ and $w_0=\frac{1}{2}$.}
\label{fig4}
\end{figure}

To realize the correlation between these quantum patterns and the
classical trajectories, note that in the minisuperspace formulation,
the cosmic evolution of the universe is modeled with the motion of a
point particle in a space with minisuperspace coordinates. In this
sense, one of the most important features in quantum cosmology is
the recovery of classical solutions from the corresponding quantum
model or, in other words, how can the WDD wave functions predict a
classical universe. In quantum cosmology, one usually constructs a
coherent wave packet with suitable asymptotic behavior in the
minisuperspace, peaking in the vicinity of the classical trajectory.
In figure \ref{fig4}, we have plotted the qualitative behavior of
the square of the wave function (\ref{AU}) with the above mentioned
Gaussian weight factor and its contour plot for typical numerical
values of the parameters. As this figure shows, while the wave
function has its dominant peaks in the vicinity of the classical
trajectories, these peaks predict a universe to come out of a
non-zero value of the scale factor. Therefore, it can be seen that
by the quantization of the model, while we are able to eliminate the
classical singularity, we are also led to a quantum pattern with a
good agreement with its classical counterpart.

\subsection{Non-flat quantum universe with zero cosmological constant: $k\neq 0$,
$\Lambda=0$}

In this case equation {\ref{AL}), the WDW equation for the scale
factor takes the form

\begin{equation}\label{AW}
\frac{d^2\psi(a)}{da^2}+4\left(-g_ca^2+g_r+\frac{g_s+w^2}{a^2}\right)\psi(a)=0,
\end{equation}in which we have set the ordering parameter as
$\beta=0$. The general solutions to this differential equation are

\begin{equation}\label{AV}
\psi(a)=C_5\,\frac{\mathrm{M}_{\mu\nu}\left(2\sqrt{g_c}\,a^2\right)}
{\sqrt{a}}\,+\,C_6\,\frac{\mathrm{W}_{\mu\nu}\left(2\sqrt{g_c}\,a^2\right)}{\sqrt{a}},
\end{equation}where $\mathrm{M}_{\mu\nu}(x)$ and $\mathrm{W}_{\mu\nu}(x)$ are
Whittaker functions with $\mu=\frac{g_r}{2\sqrt{g_c}}$ and
$\nu=\frac{i}{4}\sqrt{16(g_s+w^2)-1}$. If as in the classical case
we take $g_r=0$ (or equivalently $g_3=-2g_2$), the Whittaker
functions can be expressed in terms of the modified Bessel functions
$K_{i\nu}(z)$ and $I_{i\nu}(z)$ as follows \cite{Abra}

\begin{eqnarray}\label{AX}
\psi(a)=C_7\,\sqrt{a}\,\mathrm{K}_{\frac{i}{4}\sqrt{16(g_s+w^2)-1}}\left(\sqrt{g_c}\,a^2\right)\,
+\,C_8\,\sqrt{a}\,\mathrm{I}_{\frac{i}{4}\sqrt{16(g_s+w^2)-1}}\left(\sqrt{g_c}\,a^2\right).
\end{eqnarray} Since the wave functions must satisfy $\lim
\psi(a)_{a\rightarrow \infty}=0$, we restrict ourselves to consider
only the modified Bessel function $K_{i\nu}(z)$ as solution and so
we set $C_8=0$. The other part of the WDW equation is the equation
of the scalar field which in this case is also the same as equation
(\ref{AR}) and its solutions have already been given in relation
(\ref{AS}). Therefore, if the coefficients are selected as $C_3=1$
and $C_4=i$, the eigenfunctions of the WDW equation read

\begin{equation}\label{AY}
\Psi_w(a,\phi)=\sqrt{a\phi}\,\mathrm{K}_{\frac{i}{4}\sqrt{16(g_s+w^2)-1}}\left(\sqrt{g_c}\,a^2\right)\,
\mathrm{H}^{(1)}_{\frac{1}{m+2}}\left(\frac{4\sqrt{\lambda}w}{m+2}\phi^{\frac{m+2}{2}}\right),
\end{equation}which their superposition gives the total wave
function as

\begin{equation}\label{AZ}
\Psi(a,\phi)=\sqrt{a\phi}\int_{D'}dw\,f(w)\,\mathrm{K}_{\frac{i}{4}\sqrt{16(g_s+w^2)-1}}\left(\sqrt{g_c}\,a^2\right)\,
\mathrm{H}^{(1)}_{\frac{1}{m+2}}\left(\frac{4\sqrt{\lambda}w}{m+2}\phi^{\frac{m+2}{2}}\right),
\end{equation}where $f(w)$ is again a Gaussian-like weight factor in
the form (\ref{AU1}) and $D'$ is the domain of integration over $w$
as

\begin{eqnarray}\label{BA}
D'=\left\{
\begin{array}{ll}
(-\infty,+\infty)&g_s\geq\frac{1}{16},\\
(-\infty,-\frac{1}{4}\sqrt{1-16g_s})\cup(+\frac{1}{4}\sqrt{1-16g_s},+\infty)&g_s<\frac{1}{16}.
\end{array}
\right.
\end{eqnarray}
The results of the numerical study of this wave function are shown
in figure \ref{fig5}. The similarities and differences between
quantum and classical solutions, and the fact that the wave
functions' peaks correspond very well to the classical trajectories,
are similar to those described at the end of the previous
subsection.

\begin{figure}

\includegraphics[width=0.45\textwidth]{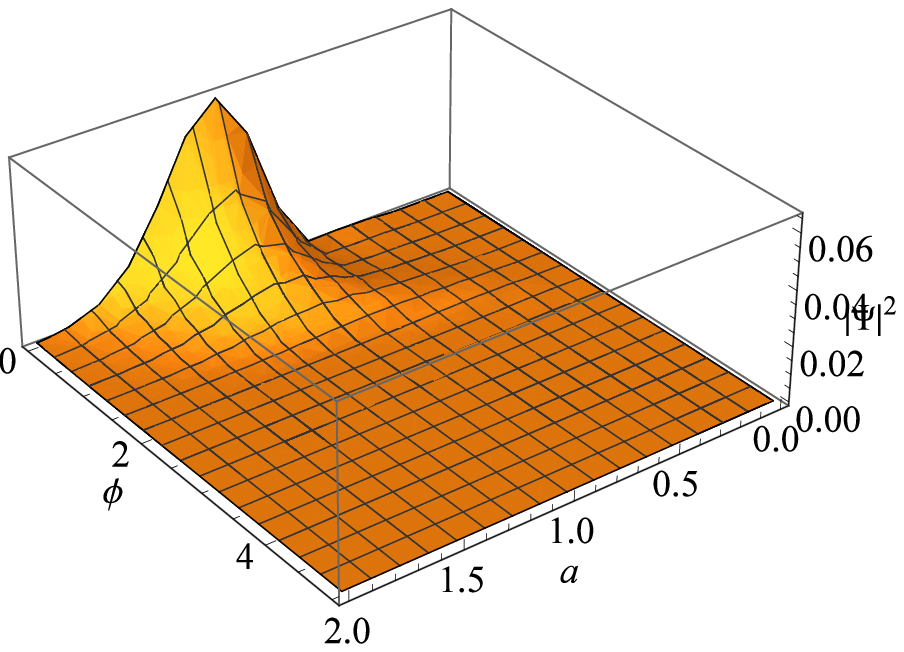}
\includegraphics[width=0.45\textwidth]{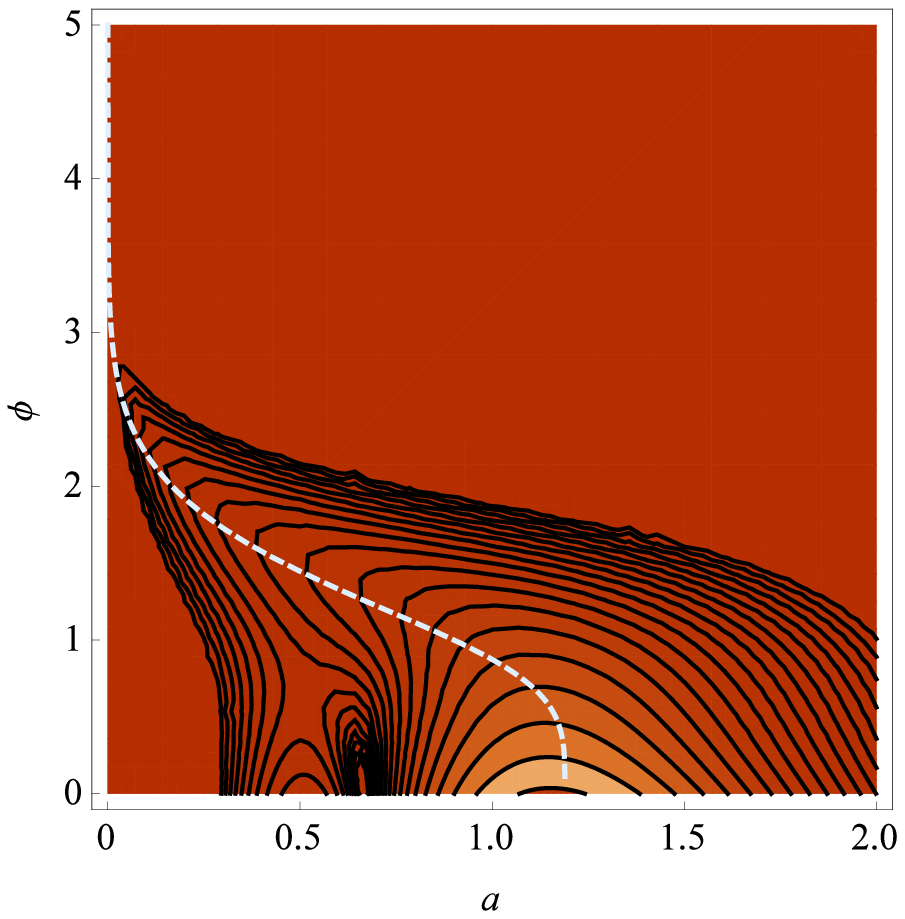}\\
\caption{The figures show the square of the wave function (left) and
its corresponding contour plot (right). Also, dashed line denotes
the classical trajectory of the system in the plane $a-\phi$. The
figures are plotted for the numerical values: $g_c=2$, $g_s=1$,
$\lambda=1$, $m=2$, $p=2$, $\sigma=\frac{1}{2}$ and $w_0=3$.}
\label{fig5}
\end{figure}

\subsection{Early quantum universe}

In this last part of the article, we study the quantum dynamics of
the model in the early times of evolution of the universe. As is
well known, the quantum behavior of the universe is more important
in this period, and it is the quantum effects in this era that
prevent the classical singularities. For small values of the scale
factor, equation (\ref{AL}) with $\beta=0$, takes the form

\begin{equation}\label{BC}
\frac{d^2\psi(a)}{da^2}+4\left(g_r+\frac{g_s+w^2}{a^2}\right)\psi(a)=0,
\end{equation}with solutions

\begin{equation}\label{BD}
\psi(a)=C_9\;\sqrt{a}\,\mathrm{J}_{\frac{i}{2}\sqrt{16(g_s+w^2)-1}}(2\sqrt{g_r}\,a)
+C_{10}\;\sqrt{a}\,\mathrm{Y}_{\frac{i}{2}\sqrt{16(g_s+w^2)-1}}(2\sqrt{g_r}\,a).
\end{equation}Since both Bessel functions satisfy the DeWitt boundary condition, both can contribute to making the wave
function, so we take the coefficients as $C_9=1$ and $C_10=i$. The
solutions for the scalar field are the same as the ones we presented
in the previous two subsections. So, the final form of the wave
function in this case is

\begin{eqnarray}\label{BE}
\Psi(a,\phi)&=&\int_{D'}dw\;f(w)\;\psi_w(a)\Phi_w(\phi)\nonumber\\
&=&\sqrt{a\phi}\;\int_{D'}dw\;f(w)\;\mathrm{H}^{(1)}_{\frac{i}{2}\sqrt{16(g_s+w^2)-1}}(2\sqrt{g_r}\,a)
\;\mathrm{J}_{\frac{1}{m+2}}\left(\frac{4\sqrt{\lambda}w}{m+2}\phi^{\frac{m+2}{2}}\right),
\end{eqnarray}where as before $f(w)$ is a Gaussian-like weight
function and the domain of integration $D'$, is given by (\ref{BA}).
The final results are shown in figure \ref{fig6}. A look at this
figure shows that the universe started its evolution from a non-zero
value for the scale factor which in turn, means that quantum effects
have eliminated the singularity of the classical model. Also, as the
figure clearly shows, the wave function has its peaks in the
vicinity of classical trajectories shown in figure \ref{fig3}, which
indicate the compatibility of classical and quantum solutions.

\begin{figure}
\includegraphics[width=0.45\textwidth]{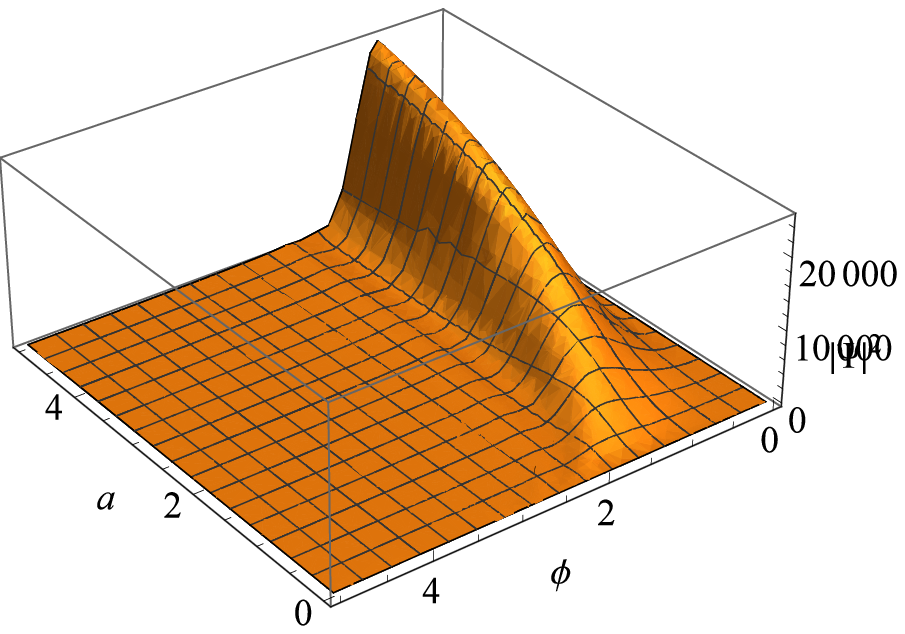}
\includegraphics[width=0.45\textwidth]{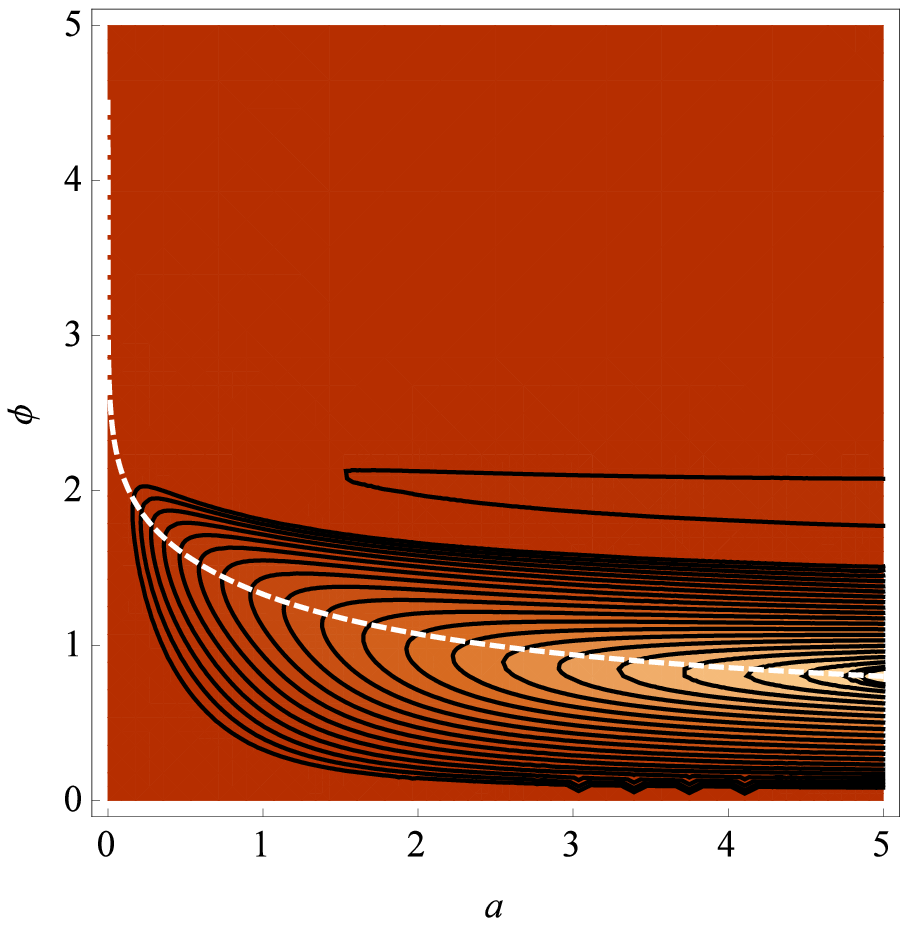}\\
\caption{The figures show the square of the wave function (left) and
its corresponding contour plot (right) for the early times of cosmic
evolution. Also, dashed line denotes the classical trajectory of the
system in the plane $a-\phi$. The figures are plotted for the
numerical values: $g_r=1$, $g_s=2$, $\lambda=1$, $m=2$, $p=0$,
$\sigma=1$ and $w_0=\frac{1}{2}$.} \label{fig6}
\end{figure}

\section{Summary}
In this paper we have studied the classical and quantum FRW
cosmology in the framework of the projectable HL theory of gravity
without the detailed balance condition. The phase space variables
turn out to correspond to the scale factor of the FRW metric and a
non-linear self-coupling scalar field with which the action of the
model is augmented. After an introductory introduction to the HL
theory, based on a dimensional analysis, we present the terms which
are allowed to be included in the potential part of the action of
this theory. This process enabled us to write the Lagrangian and
Hamiltonian of the model in terms of the minisuperspace variables
and some correction parameters coming from the HL theory.

We then studied the classical cosmology of this model and formulate
the corresponding equations within the framework of Hamiltonian
formalism. Though, in general, the classical equations did not have
exact solutions, we analyzed their behavior in the special cases of
the flat universe with cosmological constant, the non-flat universe
with vanishing cosmological constant, the early and late times of
cosmic evolution and obtained analytical expressions for the scale
factor and the scalar field in each of these cases. Another point to
note about the classical solutions is the choice of the appropriate
lapse function in each case, which actually represents the time
gauge in which that solution is obtained.  We have seen that the
classical expressions for the scale factor and scalar field exhibit
some kinds of classical singularities. These singularities are
mainly of the big-bang type for the scale factor and blowup type for
the scalar field.

The last part of the paper is devoted to the quantization of the
model described above in which we saw that the classical singular
behavior will be modified. In the quantum models, we separated the
WDW equation and showed that its eigenfunctions can be obtained in
terms of analytical functions. By an appropriate superposition of
the eigenfunctions we constructed the integral form of the wave
functions. Although it is seen that these integral expressions are
too complicated for extracting an analytical closed form for the
wave functions, employing numerical methods, we have plotted the
approximate behavior of the square of the wave functions for typical
values of the parameters. In each case, investigation of the pattern
followed by the wave functions show a non-singular behavior near the
classical singularity. In addition to singularity avoidance, we saw
that the wave functions' peaks are with a good approximation, in the
vicinity of the classical trajectories which indicate the fact that
the classical and quantum solutions are in complete agreement with
each other in the late time of cosmic evolution.

\end{document}